\documentclass{osa-article}

\journal{oe}

\usepackage{bibentry}
\newcommand{\ignore}[1]{}
\newcommand{\nobibentry}[1]{{\let\nocite\ignore\bibentry{#1}}}

\articletype{Research Article}

\newcommand{\da}[1]{\text{d}#1 }
\graphicspath{{figs/}}
\begin{document}

\title{Noise and Pulse Dynamics in Backward Stimulated Brillouin Scattering}

\author{
Oscar A. Nieves\authormark{1,*}, 
Matthew D. Arnold \authormark{1}, 
M. J. Steel\authormark{2}, 
Miko\l{a}j K. Schmidt\authormark{2} and 
Christopher G. Poulton\authormark{1}}

\address{
\authormark{1} School of Mathematical and Physical Sciences, University of Technology Sydney, 15 Broadway, Ultimo NSW 2007, Australia.\\
\authormark{2} Macquarie University Research Centre for Quantum Engineering (MQCEQ), MQ Photonics Research Centre, Department of Physics and Astronomy, Macquarie University, NSW 2109, Australia}

\email{\authormark{*}oscar.a.nievesgonzalez@student.uts.edu.au} 

\begin{abstract}
We theoretically and numerically study the effects of thermal noise on pulses in backwards Stimulated Brillouin Scattering (SBS).
Using a combination of stochastic calculus and numerical methods,
we derive a theoretical model that can be used to quantitatively
predict noise measurements. We study how the optical pulse configuration, including the input powers of the pump and Stokes fields, pulse durations and interaction time, affects the noise in the output Stokes field.
We investigate the effects on the noise of the optical loss and 
waveguide length, and we find that the signal-to-noise ratio
can be significantly improved, or reduced, for specific
combinations of waveguide properties and pulse parameters.
\end{abstract}

\section{Introduction}
Backward Stimulated Brillouin Scattering (SBS) is an opto-acoustic process that results from the coherent interaction between two counter-propagating optical fields and an acoustic wave inside an optical waveguide~\cite{eggleton2019, pant2014, brillouin1922, boyd2003,kobyakov2010}. The coupling between these fields occurs either through electrostriction or radiation pressure~\cite{rakich2012,wolff2015stimulated}, in which the electromagnetic fields induce very small strains in the material, which in turn stimulate the generation of coherent phonons via photoelasticity or via physical motion of the waveguide boundaries~\cite{rakich2012,wolff2015stimulated,sipe2016}.
This interaction has been used in various applications, including narrow-band RF and optical signal filtering~\cite{choudhary2018, jiang2018}, phase conjugation and precision spectroscopy~\cite{eggleton2019}, novel laser sources~\cite{stein2007, loh2015optica}, and in recent experiments in opto-acoustic pulse storage and memory~\cite{merklein2017}. An intrinsic property of SBS is noise: the interaction between the pump and the thermal phonon background of the waveguide generates spontaneous Stokes photons, which combine with the input signal to produce a noisy output signal. SBS noise was first studied by Boyd and Gaeta~\cite{boyd1990, gaeta1991} and Ferreira {\em et al.}~\cite{ferreira1994}, 
focusing on optical fibers. These studies assumed that the optical and acoustic
fields are all weakly guided, and furthermore examined the effects of noise 
in the quasi-continuous wave regime. Later work~\cite{kobyakov2006} extended the distributed fluctuating source model from Boyd and Gaeta to include the effect of optical loss, with the addition of an acousto-optic effective area using scalar modal fields, and was again restricted to quasi-CW pulses and weak guidance.
More recent studies have analyzed the effects of thermal noise in the context of SBS lasers and micro-ring resonators~\cite{loh2015noise,kharel2016,behunin2018, otterstrom2018}. A dynamic, fully-vectorial treatment that can be applied to 
predict noise in modern integrated SBS waveguide experiments ~\cite{ zhang2011, pant2014} is important for the further development of SBS signal-processing applications. 

In this paper, we derive a theoretical model for SBS noise for pulses in integrated optical waveguides with high-contrast materials, using fully vectorial modal fields. We focus on amplitude noise arising from the thermal phonon field, because in backwards SBS experiments this process has been observed to be significantly stronger than phase noise~\cite{pelusi2018, zarifi2020}. We explore the dependence of the noise on the waveguide and pulse properties. We find that in the regime where the pump is undepleted by the SBS process, it is possible to derive analytic results for the absolute level of noise, as well as for the signal-to-noise ratio, even for time-varying pulses of arbitrary shape. Since numerical computation of ensemble averages are expensive, an analytic model that computes the SBS noise properties of pulses is extremely useful for the design of SBS-active devices. We use this analytic model to explore the impact of pulse and waveguide properties on the noise and optical signal-to-noise ratio (OSNR), and discuss the implications of these results in SBS experiments, where it is often advantageous to minimize the Stokes noise. 

\section{Theory and formalism}
\begin{figure}[ht]
	\centering
    \includegraphics[width=1.0\textwidth]{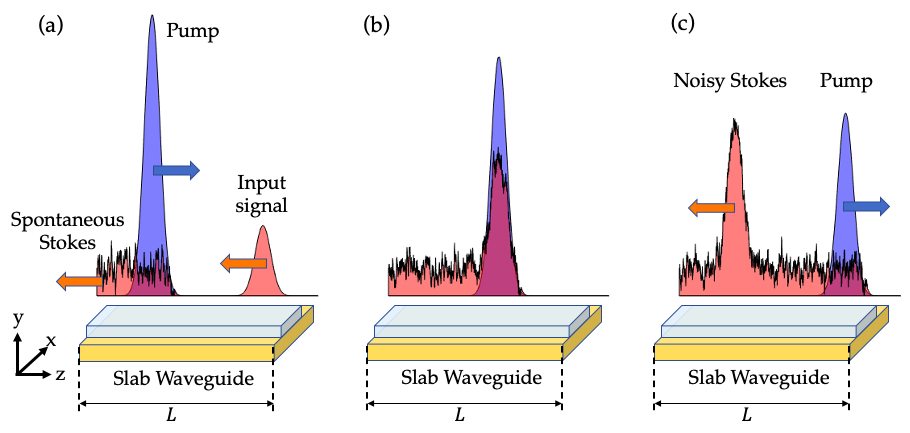}
    \caption{Propagation of two Gaussian pulses along a waveguide consisting of substrate (yellow) and core (light blue) dielectrics. $L$ denotes the total length of the waveguide medium. The input fields in (a) overlap in (b), resulting in an exchange of energy between the pump and Stokes through the acoustic field, whereby the pump's energy is depleted by the Stokes. The spontaneous Stokes noise generated by the pump interacting with the thermal phonon field combines with the coherent signal in (b), leading to a noisy output field at the Stokes frequency in (c).}
    \label{fig:Pulses_diagram}
\end{figure}

We consider backward SBS interactions in a waveguide of finite length $L$ oriented along the $z$-axis (see Fig.~\ref{fig:Pulses_diagram}): a pump pulse with angular frequency $\omega_1$ is injected into the waveguide at $z=0$ and propagates in the positive $z$-direction, while a signal pulse is injected at $z=L$ and propagates in the negative $z$-direction. The signal pulse
has frequency centred around the Brillouin Stokes frequency $\omega_2=\omega_1-\Omega$, which is down-shifted from the pump by the Brillouin shift $\Omega$, and has a spectral width that lies entirely within the Brillouin linewidth.
 When these two 
pulses meet,  the pump transfers energy to the signal field, resulting in
coherent amplification of the signal around the Brillouin frequency. At the same time, as the pump moves through the waveguide, it interacts with the thermal phonon field and generates spontaneous Stokes photons which also propagate in the negative $z$-direction. This spontaneous Stokes field combines with the coherent signal to form a noisy amplified output field centered around the Stokes frequency.

To compute the noise transferred to the signal from the thermal field, we first solve the coupled equations of the pump $\mathbf{E}_1$, Stokes $\mathbf{E}_2$  and acoustic $\mathbf{U}$ fields. We write these fields using the ansatz
\begin{align}
\label{eq:E1} \mathbf{E}_1 &= \tilde{a}_1(z,t) \, \mathbf{e}_1(x,y) \,\mathrm{e}^{i(k_1 z - \omega_1 t)} + \text{c.c.}\\
\mathbf{E}_2 &= \tilde{a}_2(z,t) \,\mathbf{e}_2(x,y) \,\mathrm{e}^{i(k_2 z - \omega_2 t)} + \text{c.c.}\\  
\label{eq:U} \mathbf{U} &= \tilde{b}(z,t)\,\mathbf{u}(x,y) \,\mathrm{e}^{i(qz - \Omega t)} + \text{c.c.},
\end{align}
where $\tilde{a}_1$, $\tilde{a}_2$ and $\tilde{b}$ are dimensionless envelope functions, $\mathbf{e}_1$ and $\mathbf{e}_2$ are the electromagnetic mode fields 
of the waveguide (with units of V/m), $\mathbf{u}$ is the waveguide displacement mode field (with units of m), and $\text{c.c.}$ denotes the complex conjugate. The parameters $k_1>0$ and $k_2<0$ represent the pump and Stokes wavenumbers respectively. The total electric field is $\mathbf{E} = \mathbf{E}_1 + \mathbf{E}_2$. For the acoustic field, $\Omega$ is the phonon angular frequency and $\Omega$ is the acoustic wavenumber. Using this ansatz along with the phase-matching conditions $k_1=k_2+q$ and $\omega_1=\omega_2+\Omega$, the governing equations for the envelope fields are~\cite{wolff2015stimulated,sipe2016}
\begin{align}
\label{eq:A1pde}\frac{\partial a_1}{\partial z} + \frac{1}{v} \frac{\partial a_1}{\partial t}  + \frac{1}{2}\alpha a_1 &= i\omega_1 Q_1 a_2 b^*,\\
\label{eq:A2pde}\frac{\partial a_2}{\partial z} - \frac{1}{v} \frac{\partial a_2}{\partial t}  - \frac{1}{2}\alpha a_2 &=  i\omega_2 Q_2 a_1 b,\\
\frac{\partial b}{\partial z} + \frac{1}{v_a} \frac{\partial b}{\partial t}  + \frac{1}{2}\alpha_{\text{ac}} b &= i\Omega Q_a a_1^* a_2 + \sqrt{\sigma} R(z,t). \label{eq:bpde}
\end{align}
In these equations, the new envelope functions $a_1$, $a_2$ and $b$ are related to the dimensionless envelope functions via the modal powers $\mathcal{P}_j$ via $a_{1,2} = \sqrt{\mathcal{P}_{1,2}} \tilde{a}_{1,2}$ and $b = \sqrt{\mathcal{P}_a} \tilde{b}$, and related to the measured optical and acoustic powers $P_{1,2}$ and $P_a$ (in Watts) via $|a_{1,2}|^2 = P_{1,2}$ and  $|b|^2 = P_a$. The boundary conditions for the pump and signal fields are applied by specifying the input values $a_1(0,t)$ and  $a_2(L,t)$ respectively. These boundary conditions can be deterministic, as is the case in our calculations, or stochastic if laser noise is incorporated into the model. The analytic results presented in this paper are valid for both cases. 

The waveguide has optical loss $\alpha$ and acoustic attenuation $\alpha_{\text{ac}}$ (in units of m$^{-1}$), with optical group velocity $v$ and acoustic group velocity $v_a$. The coefficients $Q$ represent the coupling strength of the SBS interaction, and can be calculated from the optical and acoustic modes of the waveguide~\cite{sturmberg2019}. From conservation of energy, we have $Q_2 = Q_1^*$ and $Q_a=Q_1$~\cite{wolff2015stimulated}.
 
The function $R(z,t)$ represents a complex-valued space-time white noise with zero mean $\left\langle R(z,t)\right\rangle = 0$ and auto-correlation function $\left\langle R(z,t) R^*(z',t') \right\rangle = \delta(z-z')\delta(t-t')$, where $\delta$ is the Dirac-delta function. This type of delta correlation arises in the standard treatment of noise in optical fibers~\cite{drummond2001}, and
 arises from the effective continuum of phonon modes supported by the waveguide and coupled through dissipation to the coherent excitation of the elastic field. The strength of the noise $\sigma$ gives the conversion of coherent elastic energy in the medium into thermal phonons, in the absence of any nonlinear optical effects~\cite{boyd1990}, and is given by (see Appendix A)
 \begin{equation} \label{eq:sigmadef}
\sigma = k_{\text{B}} T \alpha_{\text{ac}},
\end{equation}
where $k_B T$ is the thermal energy per phonon mode. This quantity is directly proportional to the acoustic loss in the medium $\alpha_{\text{ac}}$~\cite{pecseli2000} in accordance with the fluctuation-dissipation theorem~\cite{lima2000}. 

\subsection{Noise under the undepleted pump approximation}
An analytic solution to equations (\ref{eq:A1pde})$-$(\ref{eq:bpde}) exists under the undepleted pump approximation, which is valid for the case in which the signal power $P_2$ remains much smaller than the pump power $P_1$. We also assume that the depletion of the pump by the thermal field is negligible compared to optical loss~\cite{ferreira1994}. The equation for the pump envelope is therefore
\begin{equation}
\label{eq:A1pde_UPA}
\frac{\partial a_1}{\partial z} + \frac{1}{v} \frac{\partial a_1}{\partial t} +\frac{\alpha}{2}  a_1= 0,
\end{equation}
which, for an input field $a_p(t) = \sqrt{P_{\text{pump}}^{\text{in}}(t)}$ at $z=0$ has the solution
\begin{equation}
\label{eq:A1sol}
a_1(z,t) = a_p\left(t - \frac{z}{v} \right) \mathrm{e}^{-\alpha z/2},
\end{equation}
by the method of characteristics. Next, we simplify~\eqref{eq:bpde} by letting the term $\partial_z b \rightarrow 0$, because for experimentally relevant time scales the propagation speed of the acoustic pulses is much smaller than for optical pulses~\cite{boyd2003}. This leads to the equation for the acoustic field
\begin{equation}
\frac{1}{v_a}\frac{\partial b}{\partial t} + \frac{1}{2}\alpha_{\text{ac}} b = i\Omega Q_a a_1^* a_2 + \sqrt{\sigma} R(z,t),
\end{equation}
which can be integrated in time formally as
\begin{equation}\label{eq:bfield}
b(z,t) =  i v_a \Omega Q_a \int_{-\infty}^{t} \mathrm{e}^{\frac{1}{2}\Gamma(s-t)}a_1^*(z,s) a_2(z,s)\, \da{s} + D(z,t),
\end{equation}
where $\Gamma = v_a\alpha_{\text{ac}}$ is the decay rate of the acoustic phonons, which is related to  the Brillouin linewidth $\Delta\nu_B$ by $\Gamma = \Delta\nu_B/2\pi$, and the time-integrated noise function $D(z,t)$ is defined as
\begin{equation}
D(z,t) = v_a\sqrt{\sigma} \int_{-\infty}^{t} \mathrm{e}^{\frac{1}{2} \Gamma (s-t)} R(z,s)\, \da{s}.
\end{equation}
According to the properties of normally distributed random processes, the integral of a normal random process is also a normal random process~\cite{kubo1957, ito1951}. This means that $D(z,t)$ has zero mean $\left\langle D(z,t) \right\rangle = 0$ and auto-correlation function
\begin{equation}
\left\langle D(z,t) D^*(z',t')\right\rangle = \frac{v_a\sigma}{\alpha_{\text{ac}}}\delta(z-z') \exp\left\{-\frac{1}{2}\Gamma\left|t - t'\right|\right\}.
\end{equation}
At this stage, we can use a separation of time-scales in the integral in~\eqref{eq:bfield}, since $1/\Gamma$ is large compared to the characteristic timescales of the optical envelope fields, and this allows us to use the Markov approximation $a_{1}^{*}(z,s)a_{2}(z,s) \approx a_{1}^{*}(z,t)a_{2}(z,t) \mathrm{e}^{-i\Delta(s-t)}$~\cite{otterstrom2018}, where $\Delta$ is a detuning parameter relative to the center of the Brillouin gain profile written as $\Delta  = \omega_2 - \omega_1 +\Omega$. Next, pulling $a_{1}^{*}(z,t)a_{2}(z,t)$ outside of the first integral in (\ref{eq:bfield}) yields the approximate expression for the acoustic envelope field
\begin{equation}\label{eq:acoustic}
b(z,t) \approx i\frac{v_a\Omega Q_a}{\Gamma/2 - i\Delta}a_1^* a_2 + D(z,t).
\end{equation}
Now we can eliminate the elastic field by substituting (\ref{eq:acoustic}) into (\ref{eq:A2pde}), which yields
\begin{equation}\label{eq:A2eq}
\frac{\partial a_2}{\partial z} -\frac{1}{v}\frac{\partial a_2}{\partial t} + \frac{1}{2}\left[g(\Delta) P_1(z,t) -\alpha\right] a_2 = i \omega_2 Q_2 a_1(z,t) D(z,t),
\end{equation}
where $g(\Delta)$ is the usual SBS gain parameter in (m$^{-1}$W$^{-1}$)
\begin{equation}\label{eq:SBS_Gain}
g(\Delta) = \frac{2v_a\omega_2 \Omega |Q_2|^2}{\Gamma/2 - i\Delta}  = g_0\frac{\Gamma/2}{\Gamma/2-i\Delta},
\end{equation}
with $g_0 = g(0)$. Equation~\eqref{eq:A2eq} is then solved by making a variable transformation to the frame of reference that is co-propagating with the Stokes field~\cite{drabek2014}, as shown in Appendix B. This leads to the solution of the Stokes envelope field
\begin{multline}\label{eq:A2sol}
a_2(z,t) = a_2(L,t)G_N(z,L,t) \\
-i\omega_2 Q_2 \int_{z}^{L}G_N(z,z',t) a_1\left(z',\ t+\frac{z-z'}{v} \right) D\left(z',\ t+\frac{z-z'}{v} \right) \da{z'},
\end{multline}
where the {\em net gain function} is expressed as
\begin{equation}\label{eq:netGain}
G_N(z,z',t) = \exp\left\{\frac{1}{2} \int_{z}^{z'}\left[g(\Delta) P_1\left(\eta,\ t+\frac{z-\eta}{v}\right) - \alpha \right] \da{\eta}  \right\},
\end{equation}
for $z<z'$. The modulus squared $|G_N(z,z',t)|^2$ of this function gives the net cumulative gain in the Stokes power between $z$ and $z'$, taking into account the effect of optical losses. As the distance between $z$ and $z'$ increases, both the amplification from the pump and the optical loss increase. However, for any time $t$ in which the pump pulse $P_1(z,t)$ lies outside the domain $[z,z']$, optical loss dominates over the SBS amplification. 

We compute the total, noisy, Stokes power by considering the Stokes envelope as a sum of two contributions: a coherent signal field, which is the Stokes in the absence of any thermal noise, and a spontaneous Stokes field which contains all the contributions from the background thermal noise. This is expressed by separating the two terms in Eq.~\eqref{eq:A2sol} as $a_2(z,t) = a_2^{\text{sig}}(z,t) + a_2^{\text{spo}}(z,t)$ where
\begin{equation}
a_2^{\text{sig}}(z,t) = a_2(L,t)G_N(z,L,t),
\end{equation}
\begin{equation}\label{eq:A2spo}
a_2^{\text{spo}}(z,t) = -i \omega_2 Q_2 \int_{z}^{L} G_N(z,z',t) a_1\left(z',\ t+\frac{z-z'}{v} \right) D\left(z',\ t+\frac{z-z'}{v} \right) \da{z'}.
\end{equation}
The ensemble-average of the total Stokes power is then
\begin{equation}
\left\langle P_2(z,t)\right\rangle = \left\langle P_2^{\text{sig}}(z,t)\right\rangle + \left\langle P_2^{\text{spo}}(z,t)\right\rangle = \left\langle \left| a_2^{\text{sig}}(z,t) \right|^2 \right\rangle + \left\langle \left| a_2^{\text{spo}}(z,t) \right|^2 \right\rangle,
\end{equation}
where the relation $\left\langle a_2^{\text{sig}} a_2^{\text{spo}} \right\rangle = 0$ holds since the input laser noise in $a_2^{\text{sig}}$ and thermal noise in $a_2^{\text{spo}}$ are statistically independent from each other. The average signal power is computed by taking the ensemble average of the input field defined by the noisy boundary condition $a_2(L,t)$
\begin{equation}\label{eq:P2sig}
\left\langle P_2^{\text{sig}}(z,t)\right\rangle = \left\langle \left|a_2(L,t)\right|^2 \left|G_N(z,L,t)\right|^2 \right\rangle.
\end{equation}
Here $G_N$ is taken inside the ensemble average because it is dependent on the pump $P_1$, which may contain input laser noise. The spontaneous Stokes power is
\begin{multline}
P_2^{\text{spo}}(z,t) = \frac{\alpha_{\text{ac}} }{4} \frac{\omega_2}{\Omega} g_0 \int_{z}^{L}G_N(z,z',t) \, a_1\left(z',\ t+\frac{z-z'}{v} \right) D\left(z',\ t+\frac{z-z'}{v} \right)\da{z'}\\
 \times \int_{z}^{L} G_N^*(z,z'',t) \, a_1^*\left(z'',\ t+\frac{z-z''}{v} \right) D^*\left(z'',\ t+\frac{z-z''}{v} \right)\da{z''}.
\end{multline}
This product of integrals can be cast as a double-integral by applying the stochastic Fubini theorem~\cite{van2005coll}. A sufficient condition for this to hold is that the integrand in~\eqref{eq:A2spo} be square-integrable and finite~\cite{veraar2012}. This condition holds since $D(z,t)$ reaches a steady-state after a short time $t$~\cite{kubo1966} and both $a_1$ and $G_N$ are bounded functions~\cite{ito1951}. The spontaneous power becomes
\begin{multline}
P_2^{\text{spo}}(z,t) = \frac{\alpha_{\text{ac}} }{4} \frac{\omega_2}{\Omega} g_0 \int_{z}^{L}\int_{z}^{L} G_N(z,z',t)G_N^*(z,z'',t) \times\\
a_1\left(z',\ t+\frac{z-z'}{v} \right)a_1^*\left(z'',\ t+\frac{z-z''}{v} \right) D\left(z',\ t+\frac{z-z'}{v} \right) D^*\left(z'',\ t+\frac{z-z''}{v} \right) \da{z''}\da{z'}.
\end{multline}
The relation $\left\langle a_1(z,t) D(z,t)\right\rangle = \langle a_1(z,t)\rangle \langle D(z,t)\rangle = 0$ holds since the laser noise in $a_1$ and the thermal noise in $D$ are not coupled to each other, so they are statistically independent. Therefore, the ensemble-averaged spontaneous Stokes is 
\begin{equation}\label{eq:P2spo}
\left\langle P_2^{\text{spo}}(z,t)\right\rangle =  \frac{k_B T\Gamma}{4} \frac{\omega_2}{\Omega} g_0 \int_{z}^{L} \left\langle P_1\left(z',\ t+\frac{z-z'}{v} \right)\left|G_N(z,z',t)\right|^2 \right\rangle \da{z'}.
\end{equation}

Equation~\eqref{eq:P2spo} is one of the main results in this paper. It gives the ensemble-averaged spontaneous Stokes power, which is proportional to the pump power $P_1(z,t)$ multiplied by the net gain over the length of the waveguide. As the temperature increases, more spontaneous phonons are generated, leading to more collisions of phonons with pump photons. The annihilation of these photons then creates spontaneous Stokes photons, which amplifies the Stokes noise in the medium. The integration from $z$ to $L$ indicates that as the pump propagates through the medium, there is an accumulation of spontaneous Stokes photons propagating in the opposite direction.

Once this spontaneous power has been calculated, a useful measure to quantify the integrity of the Stokes signal detected at the waveguide output is the OSNR. This is computed by taking the ratio of the average coherent signal power to spontaneous noise power over a specified bandwidth~\cite{agrawal2012}. In time-domain, the OSNR may be calculated over an arbitrary but suitably long time period $T_0$ using the time averaged powers
\begin{equation}\label{eq:SNR}
\text{OSNR} = \frac{\int_{0}^{T_0} \left\langle P_2^{\text{sig}}(0,t) \right\rangle \da{t}}{\int_{0}^{T_0} \left\langle P_2^{\text{spo}}(0,t)\right\rangle\da{t} }.
\end{equation}
In the following calculations, using a Gaussian pulse we choose $T_0$ as four times the intensity full width at half maximum (FWHM).

\section{Results for a chalcogenide SBS chip}
\begin{figure}[htbp]
\centering
\includegraphics[width=1.0\textwidth]{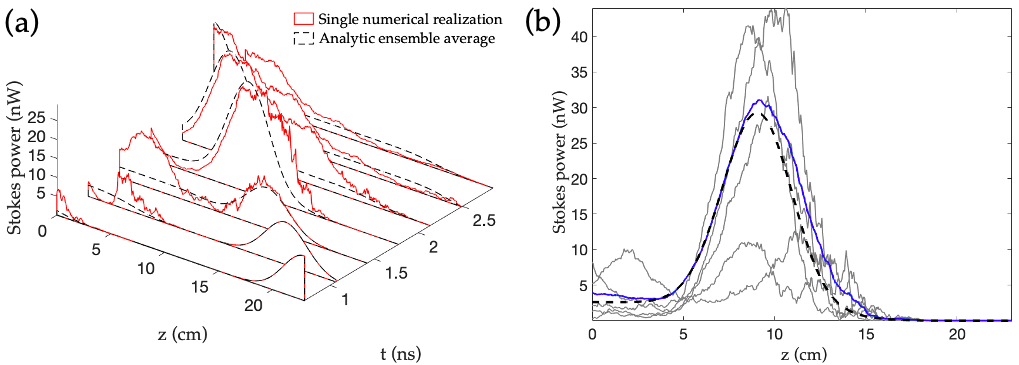}
\caption{Numerical simulation of the Stokes noise inside a 23.7 cm long waveguide. (a) Temporal evolution of the signal and spontaneous Stokes powers across the waveguide for a single run of the numerical method (red). The dashed line represents the ensemble-averaged expression for the total Stokes power being the sum of Eqs~\eqref{eq:P2sig} and~\eqref{eq:P2spo}. (b) Several independent runs of the Stokes field (gray), with the numerical ensemble average (blue) and the analytic ensemble average (black), at $t=1.8$ ns.}
\label{fig:pulses_sim}
\end{figure}

Figure~\ref{fig:pulses_sim} shows the ensemble-averaged signal power and spontaneous Stokes power computed using Eqs~\eqref{eq:P2sig} and~\eqref{eq:P2spo}. These calculations assume a dielectric waveguide at temperature 300 K of length 23.7 cm and SBS gain coefficient $g_0 = 423$ m$^{-1}$W$^{-1}$, which corresponds to the SBS chip structure used in~\cite{xie2019} at zero detuning ($\Delta = 0$). We consider a Gaussian pump pulse with peak power 110 mW and FWHM of 200 ps, and a Gaussian signal pulse with peak power 100 nW and FWHM of 200 ps. The optical pump frequency is fixed at 200 THz, while the acoustic frequency is set to 7.6 GHz, with a Brillouin linewidth of $\Omega/2\pi=30$ MHz. The linear optical loss is 0.05 dB/cm. Equations~\eqref{eq:A1pde}--\eqref{eq:bpde} are solved numerically with a split-step integration scheme that calculates the drift of the pulses first, and then uses an Euler-Mayurama step for computing the noisy nonlinear interaction between the pump and Stokes. 

For these simulations, the numerical grid consists of 600 points in $z$ and 900 points in $t$. These values were chosen, using convergence tests in the absence of noise, to ensure an  error in the gain of under $1\%$. We see in Fig.~\ref{fig:pulses_sim}(a) that as the pump propagates through the waveguide, a noisy Stokes field is generated that trails the leading edge of the pump pukse. This field combines with the input signal, with the resulting noisy Stokes field further amplified by the pump. Figure~\ref{fig:pulses_sim}(b) shows the total Stokes power as a function of $z$ at $t=1.8$ ns, where the dashed line represents the analytic solution to~\eqref{eq:P2spo} added to the coherent Stokes power in~\eqref{eq:P2sig}, and the blue line is a numerical ensemble average over 20 independent runs of the numerical method.

\begin{figure}[ht]
    \centering
    \includegraphics[width=1.0\textwidth]{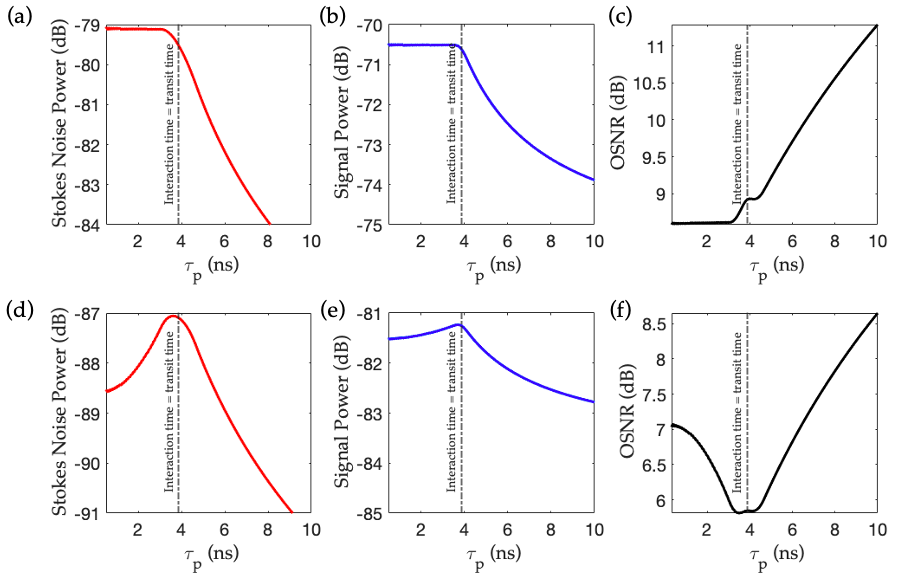}
    \caption{Time-averaged Stokes noise power, coherent signal power and OSNR for a dielectric waveguide of length $L=23.7$ cm. Plots (a)$-$(c) are for $\alpha = 0.01$ dB/cm, while plots (d)$-$(f) are for $\alpha = 1.0$ dB/cm. The signal pulse is a Gaussian with FWHM of 200 ps and peak power of 100 nW. The pump consists of a rectangular pulse with constant energy, ranging from $P_{\text{p0}} = 5$ mW ($\tau_p=10$ ns) to $P_{\text{p0}}= 5$ W ($\tau_p=10$ ps). The vertical dashed lines mark the point at which the interaction time between pump and signal becomes equal to the transit time of the signal pulse.
    }
    \label{fig:rect_pulse}
\end{figure}

\begin{figure}[ht]
    \centering
    \includegraphics[width=0.9\textwidth]{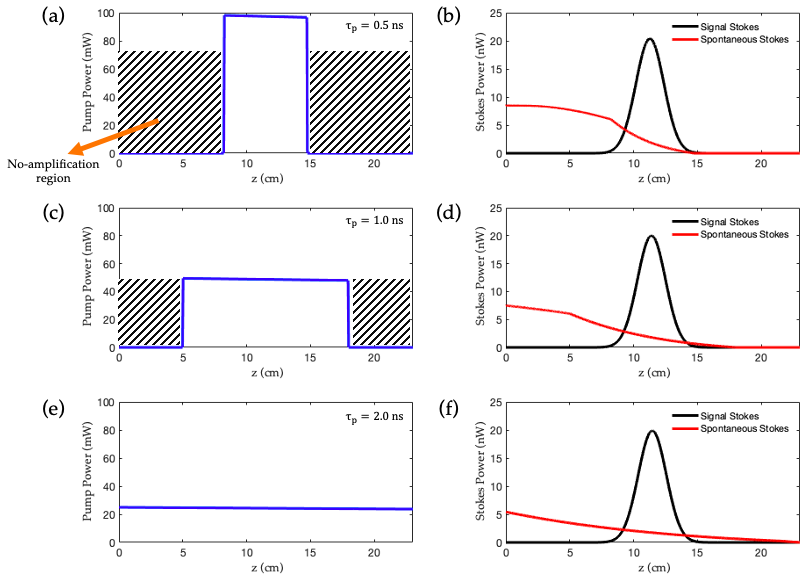}
    \caption{Snapshots of a rectangular pump pulse with varying widths and peak powers for an optical loss of $\alpha = 0.01$ dB/cm. The input pump energy is kept constant. Each row represents a snapshot of the fields inside the waveguide at fixed time $t$, for varying pump pulse widths and the corresponding signal and spontaneous Stokes powers.}
    \label{fig:pump_high1}
\end{figure}

We now use~\eqref{eq:P2spo} to study the effect on the noise of the tunable properties of the pump such as peak power and pulse duration, and how this may interact with fixed parameters such as waveguide length and optical loss. We consider a Gaussian signal pulse with FWHM of 200 ps, and a rectangular pump pulse of width $\tau_p$ that varies between 10 ps$-$10 ns. The total pulse energy of the input pump field is kept constant, while the peak power $P_{\text{p0}}$ varies with the pulse width, ranging from $P_{\text{p0}}= 5$ mW ($\tau_p=10$ ns) to $P_{\text{p0}}= 5$ W ($\tau_p=10$ ps). In all the calculations, the output Stokes field at $z=0$ is averaged over a time period of 800 ps, which is four times the signal FWHM and contains more than 99.7\% of the pulse's energy. The results for the spontaneous Stokes power, signal power and OSNR are shown in Fig.~\ref{fig:rect_pulse}. We observe in Fig.~\ref{fig:rect_pulse}(a) and (b) that the Stokes noise and signal powers remain constant for  pump widths $\tau_p \lesssim 3$ ns
before decreasing. In this low optical loss case, the near-constant values of the Stokes noise and signal power can be attributed to the constant energy in the pump pulse, as shown in Fig.~\ref{fig:pump_high1}: for short pulse-lengths, the signal and the pump interact entirely within the waveguide. As the pump pulse becomes longer than the waveguide and its peak power decreases, the Stokes field is able to interact with less of its energy inside the medium, resulting in less amplification of the Stokes. The signal and spontaneous fields fall off at different rates with respect to $\tau_p$ because the signal is proportional to the net gain squared $|G_N|^2$ via~\eqref{eq:P2sig}, while the spontaneous Stokes power is proportional to the pump power $P_1$ multiplied by $|G_N|^2$ via~\eqref{eq:P2spo}. The gain $|G_N|^2$ decreases with larger $\tau_p$ because more of the pump pulse lies outside of the interaction region, and at the same time $P_1$ decreases while its peak power is lowered for longer $\tau_p$, causing the spontaneous Stokes to decrease more rapidly than the signal. This results in an OSNR figure which increases with $\tau_p$, as shown in Fig.~\ref{fig:rect_pulse}(c).

\begin{figure}[h]
    \centering
    \includegraphics[width=0.9\textwidth]{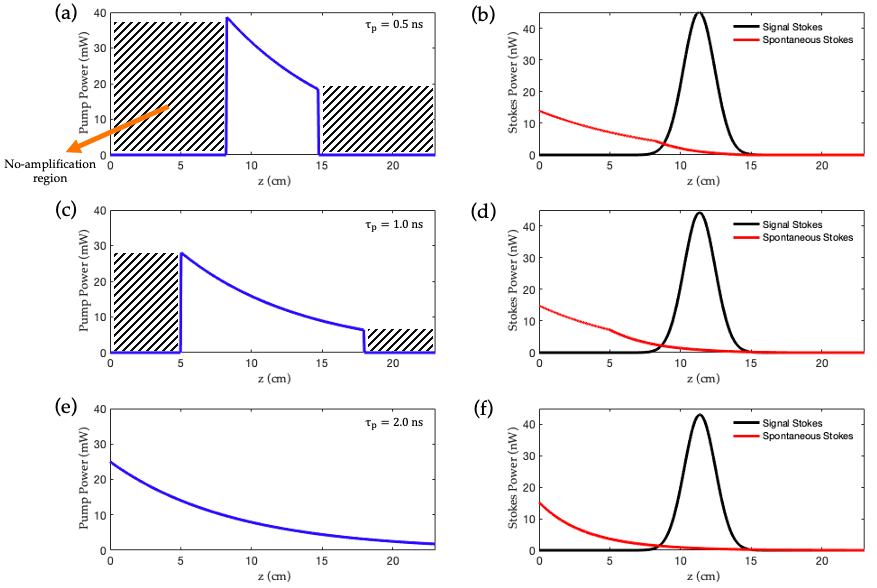}
    \caption{Snapshots of a rectangular pump pulse with varying widths and peak powers for an optical loss of $\alpha = 0.5$ dB/cm. The input pump energy is kept constant. Each row represents a snapshot of the fields inside the waveguide at fixed time $t$, for varying pump pulse widths and the corresponding signal and spontaneous Stokes powers.}
    \label{fig:pump_high2}
\end{figure}

For the high optical loss case shown in Fig.~\ref{fig:rect_pulse}(d)$-$(f), we observe a maximum in both the Stokes noise and signal powers for a critical value of $\tau_p$, before decreasing with longer pump lengths as in the lower loss case. This results in a minimum in the OSNR; this minimum occurs because the optical loss is dominant in the regions inside the waveguide that are not covered by the pump (see Fig.~\ref{fig:pump_high2}). As the pump becomes wider, the pump coverage inside the waveguide increases, compensating for the losses incurred at the uncovered regions. This results in higher amplification of the Stokes despite the peak pump power decreasing with larger $\tau_p$. However, this increased amplification vanishes once the pump pulse becomes longer than the waveguide, in which case the Stokes power decreases in a similar manner to the low-loss case.

\begin{figure}[ht]
    \centering
    \includegraphics[width=0.8\textwidth]{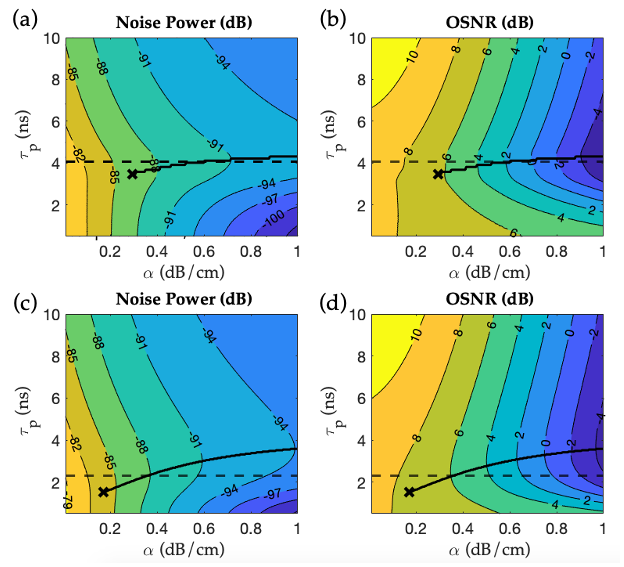}
    \caption{Contour plot of the Stokes noise power (a) and the OSNR (b) as functions of the rectangular pump width $\tau_p$ and optical loss $\alpha$, for constant energy pump pulses. Plots (c) and (d) are for Gassuan pump pulses of varying FWHM $\tau_p$. The black curve denotes where the maximum in the noise power occurs for each loss $\alpha$. The horizontal dashed line denotes when the interaction time of the pump and Stokes is equal to the transit time of the signal pulse. The cross indicates the value of $\alpha$ below which the maximum peak in the noise power vanishes.}
    \label{fig:rect_map}
\end{figure}

The emergence of the OSNR minimum in Fig.~\ref{fig:rect_pulse}(f) is a function of the optical loss $\alpha$, the pump pulse width $\tau_p$ and the waveguide length $L$. This is shown as a contour plot in Fig.~\ref{fig:rect_map}; we see that for smaller losses there is no maximum peak in the spontaneous noise power, and therefore no minimum in the OSNR. 
Below this value the Stokes power and OSNR follow the behavior of Fig.~\ref{fig:rect_pulse}(a)$-$(c), staying approximately constant for short pump pulses, then decreasing and increasing respectively for higher values of $\tau_p$.
However, once the loss reaches a critical value, the noise power develops a maximum 
at approximately the point in which the interaction time $\tau_i$ between pump and signal is equal to the transit time $L/v$. This corresponds to a minimum in the OSNR, because making $\tau_p$ longer than the transit time results in a reduced level of noise, since the spontaneous Stokes decreases more rapidly than the amplified signal.
The value $\tau_{\text{max}}$ that maximizes the noise corresponds to the pump-length when the interaction time $\tau_i$ satisfies $\tau_i = L/v$ and can be computed as follows: first, we compute an effective pulse-width $\tau_{\text{eff}}$ using the equation
\begin{equation}
    \int_{0}^{\tau_{\text{eff}}/2} P_1(0,t)\da{t} = \frac{1}{2}\tilde{f} \int_{-\infty}^{\infty} P_1(0,t)\da{t},
\end{equation}
where $0\leq \tilde{f}\leq 1$ represents the fraction of the total energy in the pump that is contained within $\tau_{\text{eff}}$. In our calculations, we set $\tilde{f}=0.95$. The interaction time between the pulses is $\tau_i=\tau_{\text{eff}}/2$ because the signal and pump counter-propagate at the same speed, so the time of overlap is halved. This gives an approximation of the value $\tau_p$ for which the maximum amount of pump energy is able to interact with the signal, before it no longer fits inside the structure.

This noise behavior is not restricted to rectangular pump pulses. In Fig.~\ref{fig:rect_map}(c) and (d), we observe the same behavior when using a Gaussian pump pulse, and again the same maximum in the Stokes noise can be observed. Here, some of the pump's energy is more localized near the central peak, while the long tails of the pulse contain a very small portion of the energy. This limits the region inside the waveguide in which the amplification from the pump is able to compensate for the optical losses, and results in a maximum Stokes noise peak that occurs at a smaller $\tau_p$ compared to the rectangular pump case. 

\begin{figure}[ht]
    \centering
    \includegraphics[width=0.85\textwidth]{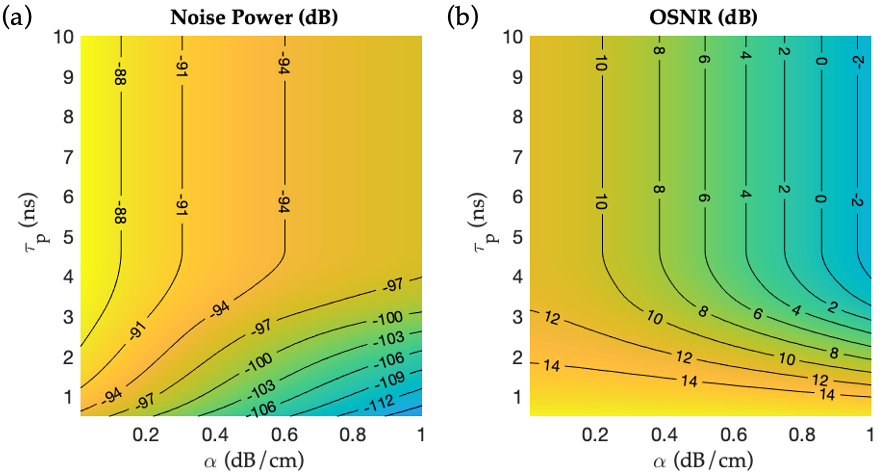}
    \caption{Stokes noise power (a) and OSNR (b) for a rectangular pump pulse using a constant peak pump power.}
    \label{fig:rect_map_c}
\end{figure}

Finally, we compute the effects of the noise and OSNR
in the case where the pump pulse has a fixed peak power. Both the spontaneous Stokes noise and signal power increase with $\tau_p$, reaching a steady state for $\tau_p$ longer than the transit time of the waveguide. This results in the OSNR decreasing until reaching a constant value for larger $\tau_p$, as shown in Fig.~\ref{fig:rect_map}(b). This occurs due to the pump pulse covering more of the waveguide as $\tau_p$ gets larger: initially there is more energy injected into the SBS medium, but as soon as the pump pulse becomes comparable in size to the waveguide; the amount of pump energy that interacts with the Stokes field reaches a constant value. This then leads to a steady-state for both the signal and spontaneous Stokes powers. 

\section{Validity of the analytic model}

We now estimate the region of validity of our analytic model, by considering an SBS chalcogenide chip with length 23.7 cm, with a Gaussian pump pulse of $P_{\text{p0}} = 300$ mW and fixed optical loss of $\alpha = 0.3$ dB/cm. The energy in both the pump and signal input fields is given by the expressions $E_{1} = \int_{-\infty}^{\infty} P_1(0,t) \da{t}$ and $\quad E_{2} = \int_{-\infty}^{\infty} P_2(L,t) \da{t}$ respectively. For each ratio of energies $E_2/E_1$ where $E_2 \leq E_1$ we compute the relative error in the OSNR using both the analytic model in~\eqref{eq:P2spo} and the numerical solver for Eqs~\eqref{eq:A1pde}$-$\eqref{eq:bpde}. 

\begin{figure}[h]
\centering
\includegraphics[width=1.0\textwidth]{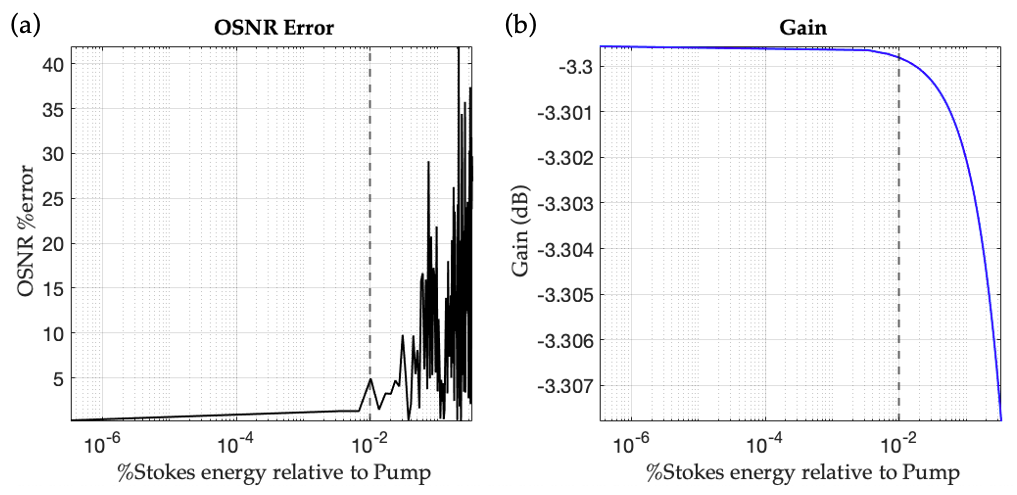}
\caption{Plots for the signal percentage error in the OSNR (a) and signal gain (b). All the plots are represented as functions of the input signal energy as a percentage of the input pump energy, over a range of signal powers between 1 nW$-$1 mW relative to a 300 mW input pump at a fixed optical attenuation of 0.3 dB/cm. The dashed vertical lines represent the point at which the near-constant gain begins to decrease, indicating that the pump is being depleted by the Stokes field.}
\label{fig:Gain_OSNR}
\end{figure}

In Fig.~\ref{fig:Gain_OSNR}(a) we see that the ONSR error remains below 2\% when the signal energy relative to the pump is small, but increases suddenly for larger signal powers. This occurs because of pump depletion, which was neglected in 
Eq.~\ref{eq:P2spo} in deriving the analytic model: as the pump
becomes depleted, some of its energy gets transferred to the spontaneous Stokes as well as to the signal, enhancing the coupling between the fields. This results in some of the spontaneous Stokes energy feeding back into the pump, making the multiplicative noise more dominant than the additive noise in Eq.~\eqref{eq:P2spo}. This then leads to the sudden increase in the OSNR error observed in Fig.~\ref{fig:Gain_OSNR}(a). The analyic model is therefore expected to yield accurate results provided the pump remains undepleted.

\section{Conclusion}
In this paper, we have developed a general mathematical model for solving the dynamical SBS equations in the presence of thermal noise. We present a simple analytic model in the undepleted pump regime whereby the amplitude noise in the Stokes field can be computed  for pulses of arbitrary shape and size, which is useful in Brillouin signal processing. The analytic results show that in the case of a constant energy pump field and lossy media, the OSNR has a minimum peak near the region in which the interaction time of the pulses matches the transit time of the signal in the waveguide. This occurs as the spontaneous Stokes noise increases towards a maximum value, which results from longer pump pulses compensating for the linear optical losses in the medium. Once the pump pulse becomes longer than the waveguide, the loss dominates again as less pump energy fits inside the SBS medium. This behaviour of the OSNR is mediated by the pump energy, pump shape, waveguide length and optical loss, making it important to choose the right parameter combination to maximize the OSNR in a specific device.

\appendix
\section{Derivation of the noise strength}
In this appendix, we find an expression for the  noise strength parameter $\sigma$ in Eq.~\eqref{eq:sigmadef} by considering the acoustic envelope field $b_j$ at a set of discrete positions along the waveguide $z_j$ separated by a small width $\Delta z$, such that $b_j=b(z_j)$, and assuming $\partial_z b$ is negligible due to the small propagation length of the phonons~\cite{boyd2003}. The thermal noise field in each slice is $\tilde{R}_j$, and we assume it to be statistically independent between any two distinct slices. Additionally, we assume a weak opto-acoustic coupling so that the $b_j$ field is mainly dominated by the thermal noise. This leads to the following equation for the acoustic envelope field in the $j$th slice in the waveguide
\begin{equation}
\label{eq:bjdef}
\frac{\mathrm{d}b_j}{\mathrm{d}t} + \frac{1}{2} \Gamma b_j = v_a \sqrt{\sigma_j} \tilde{R}_j(t),
\end{equation}
where   $\sigma_j=\sigma/\Delta z$ and $\Gamma=v_a \alpha_{\text{ac}}$. The complex-valued noise $\tilde{R}_j(t)$ is written as
\begin{equation}
\tilde{R}_j(t) = \frac{\tilde{R}_j^{(1)} (t) + i \tilde{R}_j^{(2)} (t)}{\sqrt{2}},
\end{equation}
with the properties
\begin{equation}
\langle \tilde{R}_j(t) \rangle = 0,\quad \langle \tilde{R}_j^{(1)} (t)\tilde{R}_j^{(2)}(t)\rangle = 0, \quad \langle \tilde{R}_j(t) \tilde{R}_j^* (t') \rangle = \delta(t-t').
\end{equation}
Equation~\eqref{eq:bjdef} has the solution
\begin{equation}
b_j(t) = b_j(0)\mathrm{e}^{-\frac{1}{2}\Gamma t} + v_a \sqrt{\sigma_j}\int_{0}^{t} \mathrm{e}^{-\frac{1}{2}\Gamma(t-t')} \tilde{R}_j(t') \,\mathrm{d}t',
\end{equation}
and thus the mean squared amplitude can be found via
\begin{equation}
\langle |b_j(t)|^2\rangle = |b_j(0)|^2 \mathrm{e}^{-\Gamma t} + v_a^2 \sigma_j\int_{0}^{t} \int_{0}^{t'} \mathrm{e}^{-\frac{1}{2}\Gamma(2t-t'-t'')}\langle \tilde{R}_j(t') \tilde{R}_j^*(t'')\rangle \,\mathrm{d}t'' \mathrm{d}t',
\end{equation}
and using $\langle \tilde{R}_j(t) \tilde{R}_j^* (t') \rangle = \delta(t-t')$ one obtains
\begin{align*}
\langle |b_j(t)|^2\rangle &= |b_j(0)|^2 \mathrm{e}^{-\Gamma t} + v_a^2 \sigma_j\int_{0}^{t} \int_{0}^{t'} \mathrm{e}^{-\frac{1}{2}\Gamma(2t-t'-t'')}\delta(t'-t'')\,\mathrm{d}t'' \mathrm{d}t'\\
&= |b_j(0)|^2 \mathrm{e}^{-\Gamma t} + v_a^2 \sigma_j \int_{0}^{t} \mathrm{e}^{-\Gamma(t-t')} \,\mathrm{d}t'\\
&=  |b_j(0)|^2 \mathrm{e}^{-\Gamma t} + \frac{v_a \sigma_j}{\alpha_{\text{ac}}}\left(1-\mathrm{e}^{-\Gamma t} \right),
\end{align*}
which in the limit $t\rightarrow\infty$ yields the steady-state expression
\begin{equation}
\langle |b_j(t)|^2 \rangle_{\infty} = \frac{v_a \sigma_j}{\alpha_{\text{ac}}}.
\end{equation}
This quantity represents the average acoustic power, and thus we can then relate this to the thermal energy per slice $\Delta z$ through the equipartition theorem~\cite{lima2000}, by dividing both sides by the acoustic speed $v_a$
\begin{equation}
 \frac{1}{v_a}\langle |b_j(t)|^2 \rangle_{\infty} = \frac{ \sigma_j }{ \alpha_{\text{ac}} } = \frac{k_B T}{\Delta z},
\end{equation}
which then yields
\begin{equation}
\sigma_j = \frac{k_B T \alpha_{\text{ac}} }{  \Delta z}.
\end{equation}
Then, we obtain the value for $\sigma$ based on the slice thickness $\Delta z$
\begin{equation}
\sigma = \sigma_j\Delta z = k_B T \alpha_{\text{ac}}  .
\end{equation}

\section{Full derivation of the solution to the Stokes field in the undepleted pump approximation}
In this Appendix we show the complete derivation of the analytic solution for the Stokes envelope field in~\eqref{eq:A2sol}. Starting with Eq.~\eqref{eq:A2eq} and by using the transformation of variables $\xi = z$ and $\tau = t + z/v$, one moves into a frame of reference which moves along with the Stokes field. The partial derivatives in the new coordinate system $(\xi,\tau)$ are thus
\begin{align}
\frac{\partial}{\partial t} &= \frac{\partial\xi}{\partial t} \frac{\partial}{\partial \xi}+\frac{\partial\tau}{\partial t} \frac{\partial}{\partial \tau} = \frac{\partial}{\partial \tau},\\
\frac{\partial}{\partial z} &= \frac{\partial\xi}{\partial z} \frac{\partial}{\partial \xi}+\frac{\partial\tau}{\partial z} \frac{\partial}{\partial \tau} =\frac{\partial}{\partial \xi} + \frac{1}{v}\frac{\partial}{\partial \tau}.
\end{align}
Upon substituting these derivatives into Eq.~\eqref{eq:A2pde} we obtain
\begin{equation}
\label{eq:da2dxi}
\frac{da_2}{d\xi} + \frac{1}{2}\left[g(\Delta) P_1(\xi,\tau-\xi/v) - \alpha \right]a_2 = i\omega_2Q_2a_1(\xi,\tau-\xi/v) D(\xi\tau-\xi/v),
\end{equation}
Identifying the integrating factor
\begin{equation}
I(\xi,\tau) = \exp\left\{-\frac{1}{2} \int_{\xi}^{L}\left[g(\Delta) P_1\left(\xi',\tau-\frac{\xi'}{v}\right) - \alpha \right] \da{\xi'}  \right\},
\end{equation}
leads to the equation
\begin{equation}
\int_{\xi}^{L} \frac{d}{d\xi'}\left[a_2(\xi',\tau)I(\xi',\tau) \right]\da{\xi'} = i\omega_2Q_2 \int_{\xi}^{L} I(\xi',\tau) a_1\left(\xi',\ \tau-\frac{\xi'}{v} \right) D\left(\xi',\ \tau-\frac{\xi'}{v} \right) \da{\xi'},
\end{equation}
The left-hand side of Eq.~\eqref{eq:da2dxi} is $ a_2(L,\tau)I(L,\tau)- a_2(\xi,\tau)I(\xi,\tau)$. Then, using $a_2(L,\tau)$ as the input boundary condition for the Stokes field, and introducing the cumulative net gain function
\begin{equation}
\tilde{G}_N(\xi,\xi',\tau) = \exp\left\{ \frac{1}{2} \int_{\xi}^{\xi'}\left[g(\Delta) P_1\left(\xi'',\tau-\frac{\xi''}{v}\right) - \alpha \right] \da{\xi''} \right\}.
\end{equation}
we write the solution for the Stokes envelope field at the point $\xi$ as
\begin{equation}\label{eq:Solution2}
a_2(\xi,\tau) = a_2(L,\tau)\tilde{G}_N(\xi,L,\tau)
-i \omega_2Q_2 \int_{\xi}^{L} \tilde{G}_N(\xi,\xi',\tau) a_1\left(\xi',\ \tau-\frac{\xi'}{v} \right) D\left(\xi',\ \tau-\frac{\xi'}{v} \right) \da{\xi'}.
\end{equation}
Finally, we switch back to the coordinate system $(z,t)$ by replacing $\tau$ in the above equation by $\tau = t+z/v$, yielding Eq.~\eqref{eq:A2sol}.

\section*{Disclosures}
The authors declare no conflicts of interest.

\section*{Acknowledgments}
The authors acknowledge funding from the Australian Research Council (ARC) (Discovery Projects DP160101691, DP200101893), the Macquarie University Research Fellowship Scheme (MQRF0001036) and the UTS Australian Government Research Training Program Scholarship (00099F).  Part of the numerical calculations were performed on the UTS Interactive High Performance Computing (iHPC) facility. 


\end{document}